# KilonovAE: Exploring Kilonova Spectral Features with Autoencoders

N. M. Ford[1,2], Nicholas Vieira[1,2], John J. Ruan[3], and Daryl Haggard[1,2]
[1] Department of Physics, McGill University, 3600 rue University, Montreal, Québec, H3A 2T8, Canada; nicole.ford@mail.mcgill.ca
[2] Trottier Space Institute, 3550 Rue University, Montréal, Québec, H3A 2A7, Canada
[3] Department of Physics and Astronomy, Bishop's University, 2600 rue College, Sherbrooke, Québec, J1M 1Z7, Canada
*Received 2023 May 31; revised 2023 November 6; accepted 2023 November 7; published 2024 January 18*

## Abstract

Kilonovae are likely a key site of heavy *r*-process element production in the Universe, and their optical/infrared spectra contain insights into both the properties of the ejecta and the conditions of the *r*-process. However, the event GW170817/AT2017gfo is the only kilonova so far with well-observed spectra. To understand the diversity of absorption features that might be observed in future kilonovae spectra, we use the TARDIS Monte Carlo radiative transfer code to simulate a suite of optical spectra spanning a wide range of kilonova ejecta properties and *r*-process abundance patterns. To identify the most common and prominent absorption lines, we perform dimensionality reduction using an autoencoder, and we find spectra clusters in the latent space representation using a Bayesian Gaussian Mixture model. Our synthetic kilonovae spectra commonly display strong absorption by strontium $_{38}$Sr II, yttrium $_{38}$Y II, and zirconium $_{40}$Zr I–II, with strong lanthanide contributions at low electron fractions ($Y_e \lesssim 0.25$). When a new kilonova is observed, our machine-learning framework will provide context on the dominant absorption lines and key ejecta properties, helping to determine where this event falls within the larger "zoo" of kilonovae spectra.

*Unified Astronomy Thesaurus concepts:* Neutron stars (1108); R-process (1324); Radiative transfer simulations (1967); Spectral line identification (2073); Dimensionality reduction (1943)

## 1. Introduction

Compact object mergers involving a neutron star (NS; either NS–NS or NS–black hole merger) are likely an important site of heavy element nucleosynthesis in the Universe. Neutron-rich material ejected during the merger can synthesize heavy elements via rapid neutron capture (the *r*-process; Burbidge et al. 1957; Cameron 1957). Numerical simulations have shown that these mergers can produce vastly different quantities and varieties of *r*-process elements, depending on the exact properties of the merging binary and the resulting physical conditions of the post-merger ejecta (e.g., Lattimer & Schramm 1974; Eichler et al. 1989; Freiburghaus et al. 1999; Rosswog et al. 1999; Goriely et al. 2011; Korobkin et al. 2012; Perego et al. 2014; Wanajo et al. 2014; Just et al. 2015; Kasen et al. 2017; Kawaguchi et al. 2020; Tanaka et al. 2020). To determine the contribution of the merger channel to producing the cosmic abundances of various *r*-process elements, we need observational measurements for which (and how many) elements are produced in mergers. Information on element production is encoded in the electromagnetic emission from the kilonova counterpart, which is expected to follow most mergers involving an NS (see Metzger 2019 for a review). A kilonova is a days- to weeks-long ultraviolet/optical/infrared transient produced by the decay and reprocessing of heavy radioactive elements formed in the post-merger ejecta. To date, we have confidently observed only one confirmed kilonova, from NS–NS merger GW170817 (Abbott et al. 2017). The kilonova associated with GW170817 (called AT2017gfo) was observed with photometry (Andreoni et al. 2017; Arcavi et al. 2017; Coulter et al. 2017; Cowperthwaite et al. 2017; Díaz et al. 2017; Drout et al. 2017; Evans et al. 2017; Hu et al. 2017; Kasliwal et al. 2017, 2022; Lipunov et al. 2017; Soares-Santos et al. 2017; Tanvir et al. 2017; Troja et al. 2017; Utsumi et al. 2017; Valenti et al. 2017; Villar et al. 2018) and spectroscopy (Chornock et al. 2017; Nicholl et al. 2017; Pian et al. 2017; Shappee et al. 2017; Smartt et al. 2017). Since then, significant attention has been devoted to inferring key properties of the merger ejecta by modeling the observed kilonova light curve (see, e.g., Chornock et al. 2017; Cowperthwaite et al. 2017; Drout et al. 2017; Villar et al. 2017; Waxman et al. 2018; Kawaguchi et al. 2020; Almualla et al. 2021; Ristic et al. 2022; Kedia et al. 2023) and spectra (e.g., Kasen et al. 2017; Watson et al. 2019; Gillanders et al. 2022, 2023b; Hotokezaka et al. 2023; Pognan et al. 2023; Ristic et al. 2023; Shingles et al. 2023; Vieira et al. 2023a). Building off of the rapid developments in kilonova modeling, more radiative transfer simulations are now needed to contextualize what to expect and how to interpret future observed kilonovae beyond AT2017gfo.

While simulations exploring kilonova parameter space via light curves are numerous (e.g., Radice et al. 2018; Kawaguchi et al. 2020; Barnes et al. 2021; Wollaeger et al. 2021; Setzer et al. 2023), fewer works have investigated spectral features with line identification across parameter space (e.g., Kasen et al. 2017; Wollaeger et al. 2018; Korobkin et al. 2021). The sparsity of broad kilonova spectral studies is due in part to the computational cost associated with generating a large set of spectral simulations. To our knowledge, there is no previous work that performs a study of kilonova spectra sampled from across the physical parameter space and links the dominant spectral absorption features to specific *r*-process species' transition lines. In this work, we build an expanded data set of simulated kilonova spectra to accomplish this task.

A key challenge faced by spectral studies is how to extract information and meaningful trends given the data volume and high dimensionality. A broad study of potential kilonova spectra requires thousands of individual spectra computed from multiple input parameters, with each spectrum having several thousand wavelength bins, and potentially multiple epochs of data. One emerging strategy for dealing with these challenges is to apply dimensionality reduction techniques. In particular, an unsupervised approach trained on the spectra is suitable for scenarios where the links between underlying physical properties and the emergent spectrum are not clear. Many studies have applied unsupervised machine learning for dimensionality reduction to process spectral or light curve data sets from exoplanets (e.g., Matchev et al. 2022), galaxies and active galactic nuclei (e.g., Yip et al. 2004a, 2004b; Portillo et al. 2020; Tardugno Poleo et al. 2023), X-ray binaries (Ricketts et al. 2023), supernovae and their remnants (Ishida et al. 2017; Iwasaki et al. 2019), and other applications.

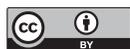

In this work, we apply unsupervised dimensionality reduction to kilonova spectra. We present the first study characterizing the dominant absorbing species and physical properties of different types of kilonova at several post-merger epochs based on their spectral features in the reduced-dimensional latent space. We test how the observed AT2017gfo spectra relate to our simulated data sets in the latent space by comparing which spectra—with identified absorption features and physical properties—they closely resemble.

Our study is presented as follows. In Section 2 we describe our methods for generating spectra, performing dimensionality reduction, and identifying clusters of spectra with similar features. In Section 3 we focus on the 1.5 days post-merger data set, reviewing the key spectral absorption signatures and their corresponding ions and dominant transition lines for each cluster. We tie these cluster-specific features to a unique combination of kilonova physical properties, and we compare AT2017gfo with the identified clusters. In Section 4 we discuss the implications of our kilonova spectra clusters, compare with other studies, and summarize some of the limitations of our analysis. We also consider how this work can be applied to future kilonova observations. We conclude in Section 5. In the Appendix we briefly extend our analysis to the data sets at 2.4 and 3.4 days post-merger.

## 2. Methods

### 2.1. Monte Carlo Radiative Transfer with TARDIS

We use the 1D Monte Carlo radiative transfer code TARDIS (Temperature And Radiative Diffusion In Supernovae; Kerzendorf & Sim 2014) to generate a diverse suite of kilonovae spectra at several epochs. TARDIS assumes a 1D, spherically symmetric ejecta undergoing homologous expansion. The expanding ejecta is configured with an inner computational "photosphere" boundary defined by the user; this photosphere is assumed to emit as a blackbody. Material below that boundary is optically thick, so the emergent spectrum is generated from the line-forming region above this boundary. Previous studies have successfully fit TARDIS spectra to AT2017gfo (Smartt et al. 2017; Watson et al. 2019; Gillanders et al. 2021, 2022; Perego et al. 2022; Vieira et al. 2023a).

A TARDIS simulation begins with the initialization of parcels of blackbody photons called photon packets. Photon packets may interact with expanding ejecta shell material as they propagate, through Thomson scattering and bound–bound interactions. Photon packets that escape the ejecta outer boundary by the end of the simulation contribute to the final synthetic spectrum. For more details on TARDIS's radiative transfer implementation, refer to Kerzendorf & Sim (2014).

The 1D spherically symmetric ejecta is described by a single-temperature, density, and abundance pattern in TARDIS. Although TARDIS can simulate multi-shell (each shell has its own density and velocity) and stratified (each shell has its own abundance pattern) ejecta, we choose to simulate ejecta as one shell in order to reduce computational expense and simplify our exploration of kilonova parameter space. We define our density $\rho$ as a power-law profile in time since merger ($t$) and ejecta velocity ($v$):

$$\rho(v, t) = \rho_0 (t/t_0)^{-3} (v/v_0)^{-\Gamma}, \quad (1)$$

where $\rho_0$, $v_0$, and $t_0$ are normalization constants, with $\rho_0$ being the density at time $t_0$ and velocity $v_0$. Assuming the ejecta is undergoing homologous expansion (valid as of $10^2$–$10^3$ s post-merger; see, e.g., Metzger et al. 2010; Kasen et al. 2013), the velocity is described by $v = r/t$ and can thus be interpreted as a radial position $r$ at a time $t$. For all our simulations, $\rho_0$ is kept as a free parameter, while $v_0$ is set to $0.1c$ and $t_0$ (not to be confused with the user input time $t$) is set to 1.4 days. We fix $\Gamma = 3$, consistent with other kilonova spectral modeling studies (Kasen et al. 2017; Tanaka et al. 2017; Watson et al. 2019; Gillanders et al. 2022; Vieira et al. 2023a), and the outer computational boundary velocity to $v_{outer} = 0.35c$, consistent with Gillanders et al. (2022) and Vieira et al. (2023a). Because TARDIS simulates ejecta above the computational photosphere boundary, our simulated density—and by extension, mass—only constrains the small portion ($\lesssim 0.01\%$–$1\%$) of the ejected material in the line-forming region. Since spectroscopic observations are most sensitive to the line-forming region, we focus our simulation work on characterizing the physical properties (e.g., density) of this region. We use the fully relativistic implementation of TARDIS (Vogl et al. 2019, 2020). We use macroatom (Lucy 2002) for handling line interactions, and we assume local thermodynamic equilibrium for handling ionization fractions and level populations. This approximation should be valid up to ∼3–5 days post-merger (Pognan et al. 2022).

We parameterize our TARDIS kilonova simulations in terms of: $\log_{10}(L_{outer}/L_\odot)$, $\log_{10}(\rho_0/\text{g cm}^{-3})$, $v_{inner}/c$, $v_{exp}/c$, $Y_e$, and $s/k_B$. $L_{outer}$ is the luminosity at the ejecta outer boundary, $\rho_0$ is the density normalization term found in Equation (1), and $v_{inner}$ is the velocity at the inner photosphere boundary. The following three parameters set the abundances in the ejecta, as described in Section 2.2: $v_{exp}$ is the expansion velocity of the ejecta immediately post-merger, $Y_e$ is the electron fraction, and $s/k_B$ is the specific entropy per nucleon.

### 2.2. Kilonova Spectral Synthesis

We generate a variety of TARDIS spectra sampled from across the kilonova physical parameter space. Table 1 lists our chosen upper and lower bounds for the physical parameters introduced in Section 2.1. Some of the parameter ranges might include values that are not completely physical. Our goal is to explore a very broad selection of potential kilonovae, acknowledging that the physical parameters are currently poorly constrained by observations.

Our framework for generating synthetic kilonova spectra derives from the methods in Vieira et al. (2023a). We perform randomized Latin Hypercube Sampling[4] within the specified physical parameter limits to generate a set of 1500 unique parameter combinations for TARDIS. Each parameter combination corresponds to a different kilonova and emergent spectrum. We then derive element abundances from our $Y_e$, $v_{\rm exp}$, and $s$ by interpolating from the abundances generated by the nuclear reaction network calculations from Wanajo (2018). To generate a spectrum based on the elements in the ejecta, we also need a list of available transition lines. We use the same measured line list compiled in Vieira et al. (2023a); the list is a combination of the Vienna Atomic Line Database (VALD; Ryabchikova et al. 2015; Pakhomov et al. 2019), and a small number of Nd II (Hasselquist et al. 2016) and Ce II lines (Cunha et al. 2017) measured from the APOGEE survey (Majewski et al. 2017). The elemental abundances, user-specified physical parameter inputs, and atomic transition line list are then fed into TARDIS to generate a suite of synthetic kilonova spectra.

**Table 1**
Bounds for Physical Parameters to Be Sampled for Generating a Training Set of Kilonova Spectra Using TARDIS

| Parameter | Simulation Range |
| --- | --- |
| $Y_e$ | 0.01–0.50 |
| $\log_{10}(L_{\rm outer}/L_\odot)$ | 6.40–8.40 |
| $\log_{10}(\rho_0/{\rm g\,cm^{-3}})$ | $-16.5$–13.5 |
| $v_{\rm inner}/c$ | 0.10–0.34 |
| $v_{\rm exp}/c$ | 0.05–0.30 |
| $s(k_B/{\rm nucleon})$ | 10–35 |

### 2.3. Training Set Generation and Preprocessing

The 1500 unique physical parameter combinations serve as input to TARDIS for generating spectra at multiple epochs. We create a training set of 1500 spectra at 1.4, 2.4, and 3.4 days post-merger (consistent with the early observed epochs for AT2017gfo spectra), for a total of 4500 spectra. We caution that TARDIS is time-independent, so kilonovae at different epochs are not physically associated in time.

We preprocess the spectra before performing any further analysis. In addition to the smoothing provided by TARDIS's virtual packet technique[5], we smooth the spectra using a Savitzky–Golay filter (Savitzky & Golay 1964) with a fifth-order polynomial and a window length of 625. The smoothed spectra are then re-scaled to a wavelength range [0,1] and normalized by their total flux in that range.

### 2.4. Dimensionality Reduction with an Autoencoder

Kilonovae spectra are complex, and the relations between different spectral features and kilonova properties are highly nonlinear. Thus, we need an empirical method for learning useful information from spectra without making any astrophysical model-dependent assumptions.

One popular approach is dimensionality reduction, which finds a low-dimensional representation of the high-dimensional data. Linear methods such as principal component analysis (PCA) are often used, but these methods struggle when the data are inherently nonlinear. Because kilonovae spectra can contain nonlinear features, we employ an autoencoder (AE) to perform nonlinear dimensionality reduction.

An AE consists of an encoder layer(s) that distills input data into a reduced-dimensional latent representation, and a decoder layer(s) which decompresses information from the latent space to reconstruct the original data. The AE is trained to find the optimal encoder–decoder neural network configuration that minimizes the reconstruction error between the original input data and the output reconstruction. A well-trained latent space representation of the data should thus pick out the most distinct features of the data (i.e., the features most important for an accurate reconstruction).

Our approach for the AE follows that of Portillo et al. (2020). The neural network architecture of our Pytorch[6]-based AE is illustrated in Figure 1. We start with an input training set with each data sample containing some number of features; in our case, a spectrum composed of 5000 binned fluxes in the wavelength range 3200–24800 Å. This high-dimensional input is passed to two encoder layers with a set of weights and biases for the neurons in each layer, and a nonlinear activation function applied to the layer. We use the Leaky Rectified Linear Unit activation function. These encoder layers are followed by the latent layer, which represents the most dimensionally reduced form of the data. The latent layer is followed by the decoder layers, which mirror the encoder layers. The output reconstruction is then given by the activation of the last layer, and it has the same dimensions as the initial data. As in Portillo et al. (2020), we use the Adam optimizer (Kingma & Ba 2014) to find the encoder–decoder layers' optimal weights and biases for minimizing the output reconstruction error. We opt for a 75:25 training–validation split, leaving 1125 of the original 1500 spectra at a single post-merger epoch available for training. We use a batch size of 32 and an initial learning rate of 0.001, which decreases by a factor of 10 if the objective function does not improve after five training epochs. We use dropout regularization in the encoder layers during training. The training runs for 200 epochs by default, with the option of early stopping if the loss function does not improve.

We train AE models with 2–6 latent dimensions, with 80 different models generated for each latent space configuration. Each model has a randomly selected number of neurons per encoder–decoder layer. The second encoder layer is restricted to be smaller than the first layer and have at least ∼10 neurons, so that it is larger than the latent layer. We assess the performance of each model by comparing the reconstructed spectra to the original data set. For each latent space configuration, we select the model with the minimum mean squared error (MSE), which represents the average reconstruction error per spectrum in the data set. These models' architectures and MSEs are provided in Table 2.

To test whether the AE is the best choice for reducing the dimensionality of our data, we compare its spectra reconstruction accuracy with a PCA and a variational AE (VAE). In Figure 2, we track the minimum MSE with 2–6 latent

---

[4] https://smt.readthedocs.io/en/latest/_src_docs/sampling_methods/lhs.html
[5] See Long & Knigge (2002) and Sim et al. (2010) for an explanation of the virtual packet technique. Kerzendorf & Sim (2014) discuss the TARDIS scheme.

[6] https://pytorch.org

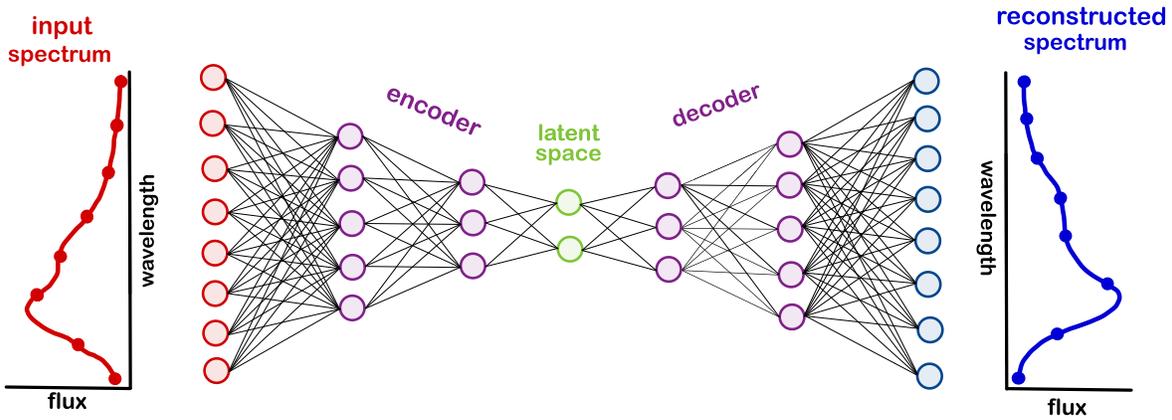

**Figure 1.** Dimensionality reduction schematic illustrating the general structure of our autoencoder (AE) with a data input layer (red), two encoder layers (purple) that learn correlations within the data and progressively reduce the dimensionality, a reduced dimension latent space (green), two decoder layers that mirror the encoder (purple), and an output reconstruction of the input data (blue).

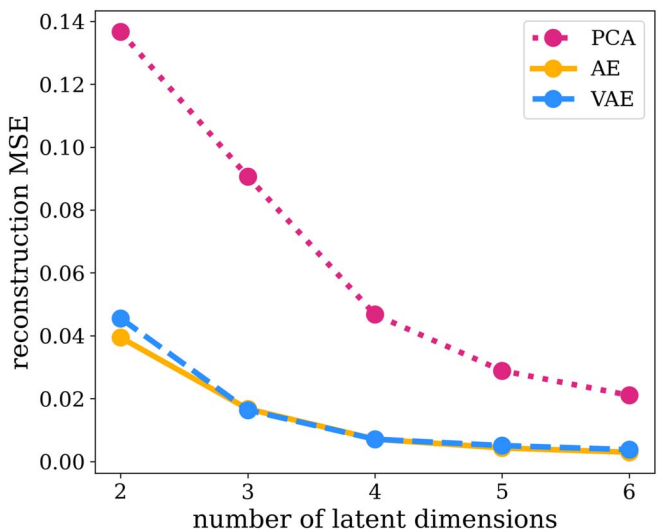

**Figure 2.** Comparison of reconstruction mean squared errors (MSE) using principal component analysis (PCA; magenta dotted line), AE (orange solid line), and variational AE (VAE; blue dashed line) dimensionality reduction for 2–6 latent space dimensions. PCA performs worst, AE and VAE perform comparably. AE and VAE MSEs plateau to a minimum value at ∼5 dimensions, with minimal improvement at 6 dimensions.

**Table 2**
Best Architectures and Spectra Reconstruction Mean Squared Error Found by Random Search for Autoencoders with 2–6 Latent Parameters for the 1.42 Days Post-merger Spectra Data Set

| Latent Dimensions | Autoencoder Architecture | MSE Loss |
|---|---|---|
| 2 | 5000-758-18-2-18-758-5000 | 0.039 |
| 3 | 5000-1842-53-3-53-1842-5000 | 0.017 |
| 4 | 5000-654-614-4-614-654-5000 | 0.007 |
| 5 | 5000-895-164-5-164-895-5000 | 0.004 |
| 6 | 5000-1028-530-6-530-1028-5000 | 0.003 |

**Note.** The architecture is written as the number of neurons in each layer, going from input on the left to the reconstructed output on the right (see Figure 1).

dimensions for each of the listed methods. For all latent space configurations, the PCA (which assumes a linear combination of components) reconstruction MSE is at least double that of the AE and VAE. The AE and VAE perform comparably, which is unsurprising considering they share similar nonlinear algorithms. A VAE's distinguishing feature is that it is designed to be generative, meaning that new spectra not in the training set could be sampled from the latent space. We opt for the regular AE in our analysis because we do not need to produce any new spectra beyond the training set.

To identify the optimal number of latent space dimensions, we look for the dimensionality at which the MSE begins to plateau. This occurs at around 5 latent space dimensions for our data (see Figure 2), and so we select the minimum MSE models with 5 latent space dimensions for further analysis. Two example spectra reconstructions using different AE latent space dimensions are shown in Figure 3. For spectra with no prominent absorption features, the AE can reproduce the original shape with just 2 latent dimensions. For spectra with more complex shapes, however, 5–6 dimensions produce the best reconstruction.

The 5D latent space reveals structures and groupings of spectra with distinct physical properties. Figure 4 shows 2D slices of the latent space colored by $Y_e$. In the latent space, the spectra are distributed with one long and narrow branch, and several "nodes" extending from the main branch (easiest to see in Figure 4's top left panel). The long branch has high $Y_e$, as does at least one of the smaller nodes, while other nodes have much lower $Y_e$. This suggests there may be some groupings of spectra with similar spectral features *and* similar underlying physical properties. We can further quantify distinct groupings of spectra using a clustering algorithm.

### 2.5. Clustering

In the low-dimensional latent space representation of the data, we can apply a clustering algorithm to identify groupings of spectra that share similar features. Clustering algorithms are unsupervised and should pick up on how certain spectra are grouped together without knowledge of the underlying physical parameters. There are many types of clustering algorithms (see Yang et al. 2022 for a recent review of clustering on spectral data); in this work we consider soft, inductive clustering algorithms. A soft algorithm is one that can assign a probability of each training set spectrum belonging to any one of the identified clusters. This is useful when certain data fall at the

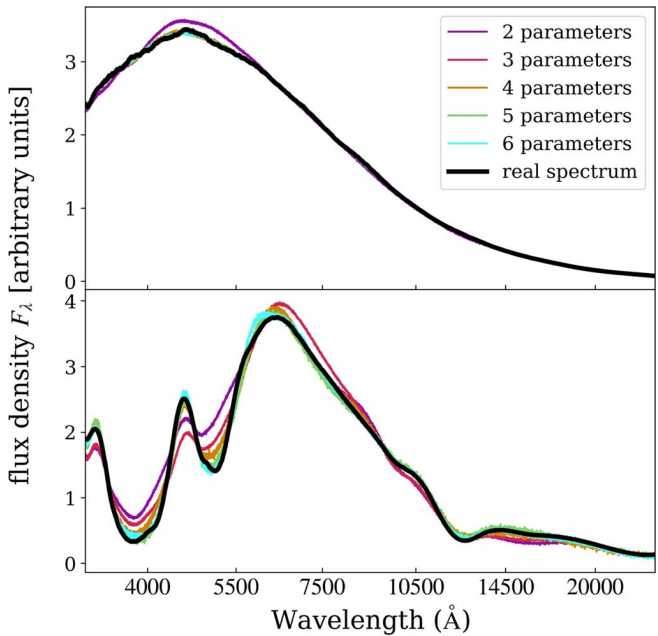

**Figure 3.** Example reconstructions of two spectra from the 1.4 days post-merger training set using different numbers of AE latent space dimensions. (top panel) The spectrum resembles a simple blackbody with no absorption features, and anything higher than two latent parameters (green curve) creates a close reconstruction of the spectrum (black line). (bottom panel) Reconstructions using 5 dimensions (red) or 6 dimensions (cyan) most closely resemble the real spectrum, particularly near the primary absorption features (∼4000 and ∼5000 Å) and the spectrum's peak (∼6000 Å).

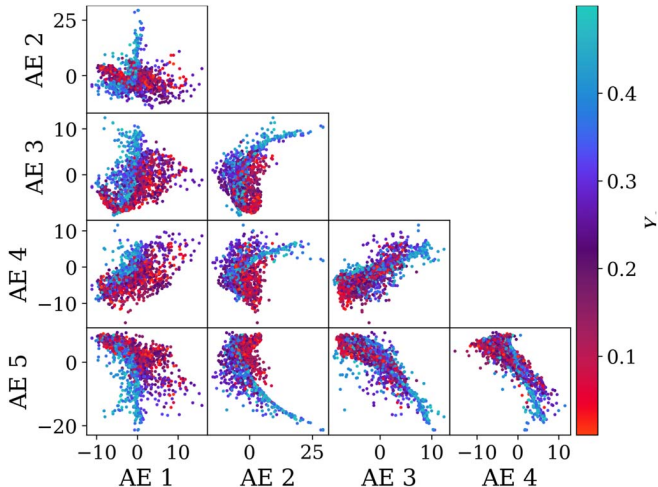

**Figure 4.** Projections of the 5D latent space representation. Each plot represents a projection comparing two latent components, and each point corresponds to one of the training set spectra. The points are colored by the electron fraction $Y_e$. There are noticeable clusters of points that share similar $Y_e$.

boundaries between clusters and cannot be assigned a confident membership, for example because they possess features associated with multiple clusters. An inductive algorithm can take any new spectrum not in the original training set and assign it a cluster membership probability based on the clusters it already found. We try several soft clustering algorithms including sklearn's Gaussian Mixture Model (GMM) with a Bayesian Information Criterion (BIC) score,[7] Bayesian

---

[7] https://scikit-learn.org/stable/modules/generated/sklearn.mixture.GaussianMixture.html#sklearn.mixture.GaussianMixture.bic

Gaussian Mixture Model (BGMM),[8] SciKit-Fuzzy's[9] "fuzzy" k-means (Ross 2010), and HDBSCAN[10] from Campello et al. 2013 (with additional dimensionality reduction using UMAP[11] from McInnes et al. 2018). We find that all methods identify similar clusters in the latent space. The BGMM has the advantage of semi-automatically identifying the "optimal" number of clusters, so we select this method for our analysis.

A BGMM is an unsupervised density estimation algorithm that assigns a combination of multivariate Gaussian probability distributions to model a data set. Each multivariate Gaussian has its own mean ($\mu$), covariance matrix ($\Sigma$), and assigned weight ($W$), which together represent a particular grouping of data within the data set. For a given number of clusters, the algorithm estimates the probability of each data point belonging to each cluster. The BGMM then updates each cluster's Gaussian $\mu$, $\Sigma$, and $W$ parameters to maximize the likelihood of the data given the estimated membership probabilities. The algorithm iterates through these two steps until convergence is reached, when the average gain on the likelihood passes below a specified threshold. sklearn's BGMM implementation takes a user-specified number of clusters as input, and then implicitly identifies the optimal number of clusters by assigning small $W$ to unnecessary "extra" clusters.

We run the BGMM on each single-epoch data set's 5D latent space representation generated from the AE model with the minimum MSE. By visual inspection, we infer that there are <13 clusters in the latent space data set, so we select this as the input cluster value to enable the BGMM to eliminate unnecessary clusters. We select a weight concentration prior of 1 and mean precision prior of 0.6; the weight concentration prior influences how many clusters receive significant weights, and the mean precision prior influences how much the identified clusters overlap. We allow the BGMM to iterate 1000 times to guarantee convergence. The convergence threshold is set to the default value of 0.001, and the algorithm usually converges in <200 iterations. Because the BGMM assigns a weight [0,1] to each cluster, it is up to the user what weight to consider significant. In practice, unless the clusters are very well separated, the BGMM may not confidently assign unnecessary clusters with precisely 0 weight. Based on trial and error of different weight cutoffs, we only include clusters with weights >0.090. If a lower cutoff is used, the additional clusters begin to resemble those already identified.

The BGMM identifies five to seven latent space clusters that capture the most distinct groupings of spectra in our data set at each epoch. For the remainder of our analysis, we present results for six identified clusters, since this is in the middle of the BGMM's "optimal" cluster range.

## 3. Results

We aim to link the spectral absorption features, the dominant absorbing species, and the physical properties of the underlying kilonovae belonging to each latent space cluster. We perform this analysis for each of three post-merger epochs, as described in Section 2.3. We find the latent space representation of AT2017gfo's spectrum at each epoch and compare it with our simulated data set to determine what sorts of cluster(s) and particular spectra it most closely resembles. For the bulk of our analysis we focus on the 1.4 days post-merger spectra, with deeper discussion of the 2.4 and 3.4 days epochs provided in the Appendix.

---

[8] https://scikit-learn.org/stable/modules/mixture.html#variational-bayesian-Gaussian-mixture
[9] https://pythonhosted.org/scikit-fuzzy/
[10] https://hdbscan.readthedocs.io/en/latest/index.html
[11] https://umap-learn.readthedocs.io/en/latest/

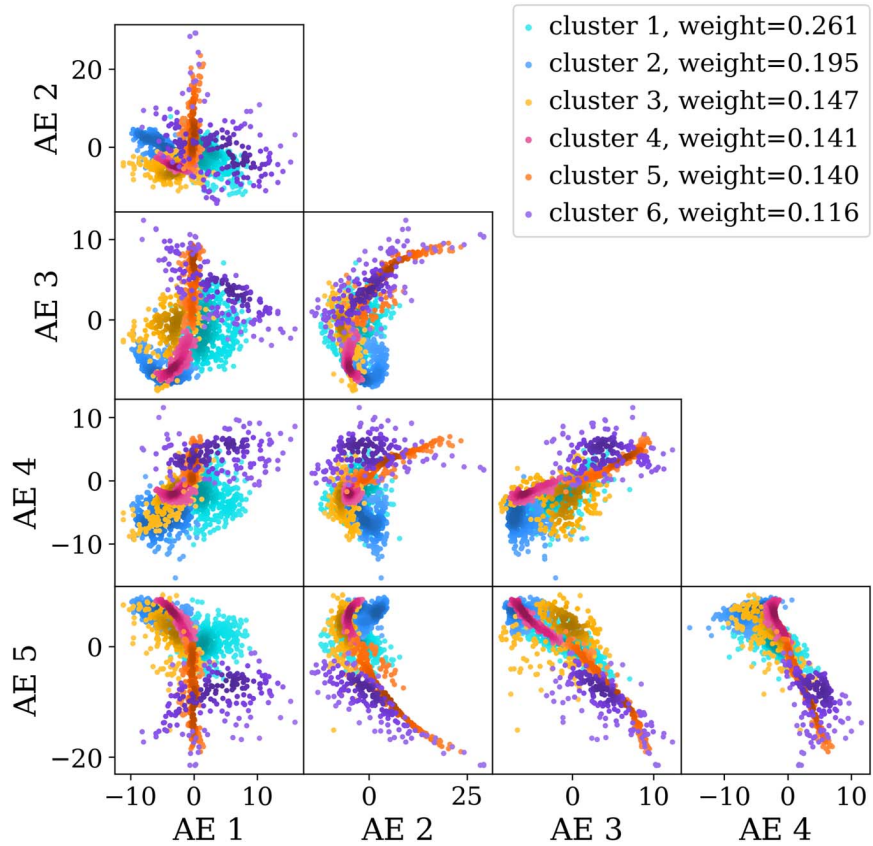

**Figure 5.** Projections of latent space components 1–5. Each plot represents a projection of the 5D latent space comparing two components, and each point corresponds to one of the 1.4 days post-merger training set spectra. The points are colored by their cluster membership. The clusters are sorted in order of descending weight. There are six clusters identified in the 1.4 days post-merger training set.

### 3.1. Cluster Membership

Each latent space spectrum is assigned membership to the cluster that it most confidently belongs to according to the BGMM. This membership probability essentially measures how similar each spectrum's shape is to each cluster mean; the closer it is to the mean spectrum, the higher the probability of it belonging to that cluster. The latent space spectra colored by cluster membership are shown in Figure 5; we list the clusters in order of descending weight. The cluster weight can be thought of as the probability of any randomly sampled point belonging to a given cluster. There is significant overlap between the clusters, and the clusters appear most distinct in projections involving the first AE latent component. The density-based BGMM clustering algorithm identifies clusters that correspond to concentrations of points in the main branch and in the nodes discussed in Section 2.4. Most data points have ≳70% confidence of belonging to a particular cluster.

Spectra belonging to each identified cluster should display distinct absorption features. To pinpoint the features that characterize each cluster, we select the subset of latent space data points with ≳99% confidence of belonging to one cluster, and compare their spectra. The spectra belonging to each cluster are shown in Figure 6.

In Cluster 1, the spectra show overlapping absorption features at many wavelengths in the 4000–14500 Å range. Cluster 2 appears redder (peaked at longer wavelengths) than most other clusters, with prominent absorption features at ∼8000 and 12 000 Å. Cluster 3 has two subgroups, one with a strong absorption feature at 8000 Å and another with a similar feature at 6500 Å.

Clusters 4 and 5 resemble featureless blackbodies, with Cluster 4 being redder and 5 being bluer. These two clusters appear in the latent space as a long and narrow branch (easiest to see in Figure 5's upper and lower left panels, with the magenta and orange clusters). We interpret this as the track of a blackbody with increasing luminosity and temperature shifting to bluer wavelengths as we move from Cluster 4 to 5.

Finally, Cluster 6 has a pronounced absorption feature at 4000 Å. There are some spectra classified in Cluster 6 which resemble a bright, blue-peaked blackbody like Cluster 5; it is likely these were misclassified because they lie in the difficult-to-distinguish overlap region between Clusters 5 and 6 (see Figure 5, orange and purple clusters).

### 3.2. Cluster Mean Spectrum

We calculate each cluster's latent space mean and translate this into a mean spectrum for the cluster. To avoid outlier spectra skewing the results, we calculate latent space means using only the spectra confidently (≳99%) associated with each cluster. This latent space mean can then be reconstructed through the trained decoder into a mean spectrum for the cluster (black lines in Figure 6). Because these mean spectra reconstructions were not part of the original training set, the AE cannot directly infer the underlying physical parameters or species dominating their absorption features. To characterize these properties, we use the Mahalanobis distance (Mahalanobis 1936), defined as

$$\sqrt{(u - l) V^{-1} (u - l)^T},\qquad(2)$$

where $u$ and $l$ are arrays corresponding to the positions of each training set spectrum and cluster mean, respectively, in the 5D latent space. $V$ is the covariance matrix of the points belonging to each cluster. We find the training set data point with the minimum Mahalanobis distance to each cluster mean. This point represents the spectrum which most closely resembles the cluster mean spectrum. Because this spectrum is from the simulated training set, we know its physical parameters and can investigate its element absorption information stored in TARDIS as a proxy for the cluster mean.

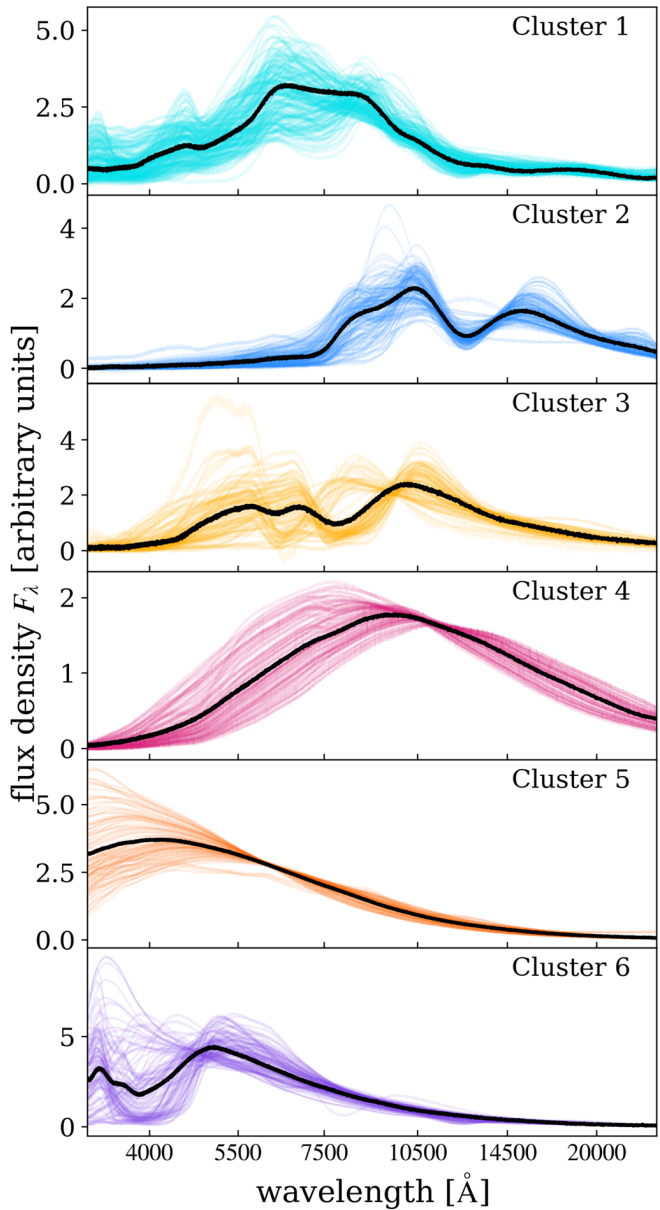

**Figure 6.** The 1.4 days post-merger spectra identified with at least 99.5% confidence to belong to one cluster, colored by cluster. The black line represents the mean spectrum for that cluster. Each cluster displays a distinct spectral shape.

### 3.2.1. Spectral Element Decomposition

We visualize the ion absorption and emission for the simulated spectrum closest to each cluster's mean using TARDIS's Spectral element DEComposition (SDEC). SDEC plots—shown for our data in Figure 7—provide a breakdown of the relative contributions of the different matter–radiation interaction modes to the output spectrum. Importantly, we can also use SDEC to distinguish which elements or ions were involved in the last interaction of the escaping photon packets in each wavelength bin. Emission is displayed above the flux density $F_\lambda = 0$ line, and includes contributions from photon packets that escape without any interactions, electron scattering, and atomic transitions. Absorption is shown below the $F_\lambda = 0$ line, with the absorbed flux indicated by the depth of the line at a given wavelength.

The top four to five species that contribute the greatest absorption are marked in different colors. Certain spectra appear to have large contributions from other species (marked in light gray), but this is mostly due to Monte Carlo noise. Both absorption and emission combine to produce the output spectrum, which traces the top of Figure 7. We also show a smoothed version of the output spectra (to minimize the Monte Carlo noise) in the left column of Figure 7. Key absorption features and corresponding elements are marked in different colors in the right column of Figure 7. The dominant transition lines contributing to these absorption features are delineated with small dashed lines in the left column.

Each cluster has distinct absorption features that can be traced to several dominant ion transition lines (or lack thereof). Cluster 1's spectrum (Figure 7, first row) has several lanthanide absorption features, with the top four contributing species being terbium $_{65}$Tb III, neodymium $_{60}$Nd III, praseodymium $_{59}$Pr III, as well as a substantial contribution from cerium $_{58}$Ce III at long wavelengths (>10 000 Å). There is also a small strontium $_{38}$Sr II feature at ∼8000 Å.

Cluster 2's spectrum displays a prominent Ce II absorption feature at ∼12 000 Å, which we attribute to three Ce II lines measured in Cunha et al. (2017). It also has prominent Sr II, lanthanum $_{57}$La II and zirconium $_{40}$Zr I absorption at ∼8000 Å.

Cluster 3 shows a very distinct Sr II absorption feature at the same location as in Clusters 1 and 2; it is produced by the Sr II 4d–5p triplet identified in Watson et al. (2019) for observed kilonova AT2017gfo. Yttrium Y II and calcium $_{20}$Ca II produce a small absorption feature at ∼6500 Å, but not all spectra in Cluster 3 display this feature (see Figure 6). A series of Y II lines dominate the short wavelength absorption, with a small contribution from Zr II, Sr II, and Ca II.

If photon packets emerge with essentially no atomic interactions, the output spectrum resembles a blackbody; this is evident in Clusters 4 and 5. The main difference between Clusters 4 and 5 is in the location and strength of the spectrum peak. Cluster 5 also has some short-wavelength absorption from a combination of Sr II/Zr II/ Y III/chromium $_{24}$Cr II.

Cluster 6 resembles Cluster 5 with two additional absorption features. Short-wavelength absorption is dominated by ruthenium $_{44}$Ru II, with small contributions from molybdenum

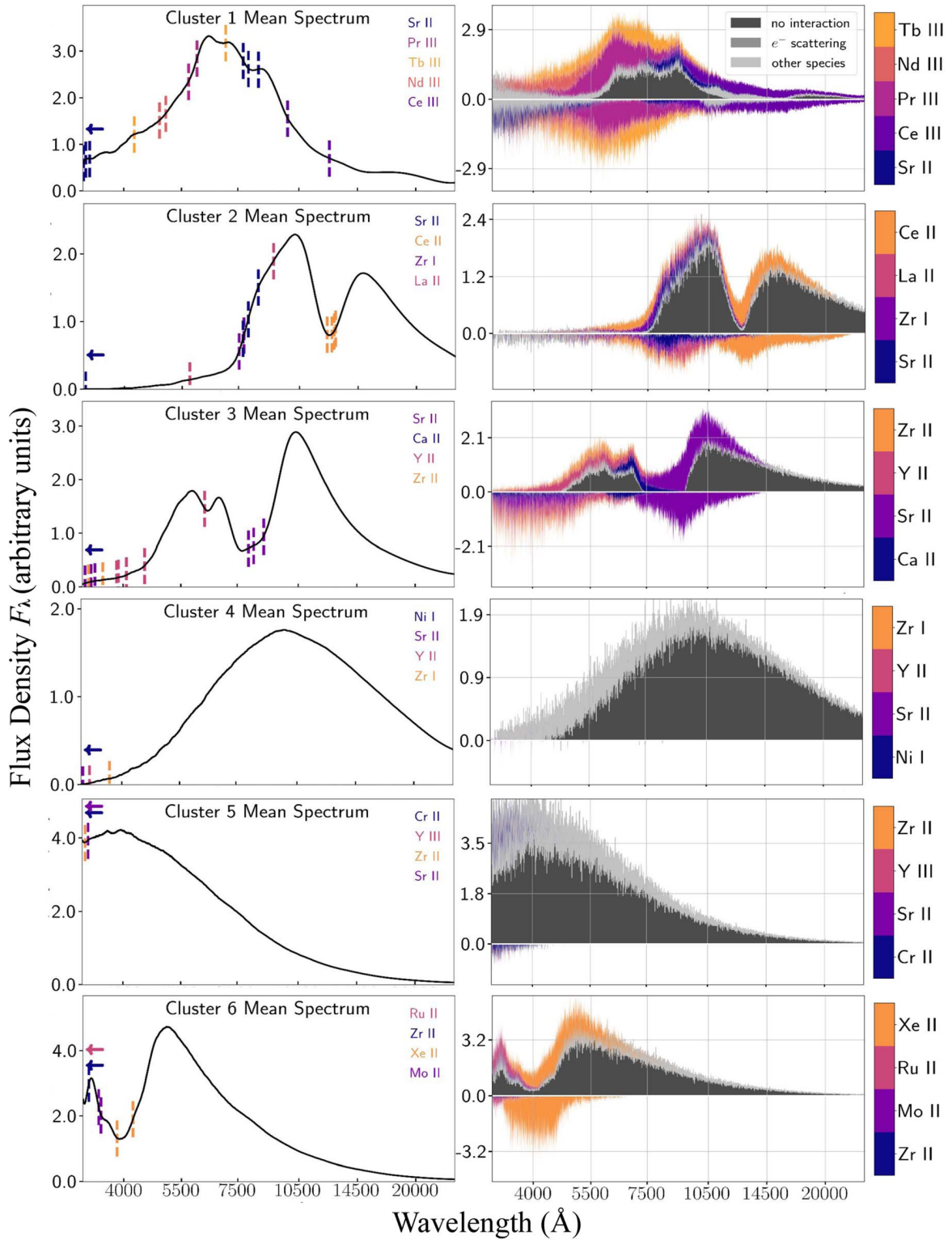

**Figure 7.** TARDIS Spectral element DEComposition (SDEC) plots for the spectrum nearest to each cluster's latent mean, as measured by Mahalanobis distance. Right column: in color are the absorption (below 0) and emission (above 0) for the top four to five contributing elements sorted by atomic number. Left column: vertical dashes mark the dominant absorption lines at their Doppler-shifted wavelengths; line centers beyond the edge of the plot are denoted with an arrow. An element may be assigned a different color based on its atomic number relative to the other plotted elements for a given cluster. Some clusters (e.g., Cluster 1), display strong lanthanide absorption, while others only show absorption from lighter $r$-process elements.

$_{42}$Mo II and Zr II. There is a strong xenon $_{54}$Xe II doublet absorption feature at ∼4000 Å.

In later epochs, spectra clusters generally display similar absorption features and physical properties to the 1.5 days post-merger data. They show some enhanced lanthanide features (see Appendix Figures 10 and 11), particularly from Ce. More detailed analyses of the 2.5 and 3.5 days post-merger data sets are provided in the Appendix.

Table 3
Physical Parameters of the 1.4 Days Post-merger Training Set Spectra Identified as Closest (Minimum Mahalanobis Distance) to Each Cluster Mean

| Cluster | Weight | $Y_e$ | $\log_{10}(L_{\rm outer}/L_\odot)$ | $\log_{10}(\rho_0/{\rm g\,cm^{-3}})$ | $v_{\rm inner}/c$ | $v_{\rm exp}/c$ | $s/k_{\rm B}$ |
|---|---|---|---|---|---|---|---|
| 1 | 0.265 | 0.205 | 8.042 | −13.809 | 0.248 | 0.244 | 10.4 |
| 2 | 0.195 | 0.206 | 6.627 | −13.999 | 0.265 | 0.253 | 12.9 |
| 3 | 0.148 | 0.288 | 7.093 | −14.349 | 0.165 | 0.087 | 21.4 |
| 4 | 0.141 | 0.351 | 6.843 | −15.963 | 0.297 | 0.262 | 17.7 |
| 5 | 0.138 | 0.372 | 8.183 | −14.947 | 0.205 | 0.078 | 24.7 |
| 6 | 0.144 | 0.324 | 7.947 | −15.563 | 0.164 | 0.165 | 32.0 |

### 3.3. Cluster Physical Parameters

In addition to the spectral features, we can characterize each cluster by its associated physical parameters. Each simulated spectrum—including the spectrum closest to each cluster's mean—has a set of physical parameters used to generate it. In Table 3 we summarize the simulation parameters for each cluster mean spectrum. The key point from the table is that the clusters not only display distinct absorption features, but also have very different physical parameters. This implies that the observed absorption features in a spectrum trace key physical parameters of the underlying kilonova. The precise values given in this table should be interpreted cautiously, since the simulated spectra do not provide an exact representation of the cluster mean. Likewise, the physical parameters corresponding to each of these simulated spectra do not convey the range of physical parameters associated with the cluster. To better understand the broad trends in physical parameters for each identified cluster, we can explore where all the spectra belonging to a certain cluster fall in the physical parameter space.

In Figure 8, we create a series of projections comparing all the physical parameters used to generate the simulated spectra. We consider the same subset of spectra confidently associated with each cluster, as discussed in Section 3.1. $Y_e$, $L_{\rm outer}$, $\rho_0$, and $v_{\rm inner}/c$ show clear concentrations for certain clusters; these concentrations are most apparent in the first column of projections. For example, Cluster 1 is characterized by low $Y_e$ and high $L_{\rm outer}$, and Cluster 6 is very concentrated toward medium $Y_e$ and high $L_{\rm outer}$. Certain clusters show much more constrained $\rho_0$ and $v_{\rm inner}/c$ ranges than others. For example, in Figure 8's upper left two panels, Cluster 4 is much more concentrated in $\rho_0$ and $v_{\rm inner}/c$ compared to Clusters 1 and 6. This indicates that $\rho_0$ and $v_{\rm inner}/c$ may have more or less of a role in shaping the dominant spectral features, depending on the overall physical configuration of the kilonova. Interestingly, in the $v_{\rm inner}/c$ versus $L_{\rm outer}$ projection, several of the clusters are diagonally spread instead of just round (see Clusters 3, 4, and 6); this is presumably due to $v_{\rm inner}/c$ scaling with $L_{\rm outer}$. Lower $v_{\rm inner}/c$ effectively corresponds to a smaller inner boundary radius for the simulation, which widens the shell of ejecta that photons pass through and strengthens the absorption, thereby decreasing $L_{\rm outer}$. No clusters show notable trends in the $v_{\rm exp}/c$ or $s/k_{\rm B}$ parameters, which indicates that the spectral appearance is generally insensitive to these parameters for our simulations (see Lippuner & Roberts 2015 for further discussion of these parameters).

### 3.4. AT2017gfo in the Latent Space

Spectra of AT2017gfo display absorption features from $r$-process elements. Initial analysis of AT2017gfo by Smartt et al. (2017) attributed a prominent absorption feature near ∼8000 Å to a combination of Cs and Te. Several studies (e.g., Watson et al. 2019; Domoto et al. 2021; Perego et al. 2022) have updated this identification and instead attribute the absorption feature to Sr. Other recent works have both confirmed the Sr feature and additionally identified long-wavelength absorption (≲5000 Å) contributions from elements such as Y and Zr (Domoto et al. 2022; Gillanders et al. 2022; Vieira et al. 2023a). At 2.4 to 3.4 days post-merger, there is also evidence for lanthanides such as Ce, Nd, La and europium $_{63}$Eu at short wavelengths (Domoto et al. 2022; Gillanders et al. 2022).

We can test the efficacy of our dimensionality reduction and clustering approach using AT2017gfo. The AT2017gfo spectra were obtained using VLT/X-shooter (Pian et al. 2017; Smartt et al. 2017).[12] Gaps in the spectra are due to a combination of poor sensitivity in the spectrograph and atmospheric absorption. To apply our trained AE to these spectra, we first fill the data gaps using an iterative PCA approach[13] similar to Portillo et al. (2020). We input the filled spectrum at each epoch into our trained AE to reduce it into the latent space. Then, our trained BGMM clustering algorithm uses the new spectrum's latent space position to determine its cluster membership probabilities. To interpret AT2017gfo's absorption features and physical properties, we apply the same method as in Section 3.2 and find a training set spectrum that approximates the new spectrum. We consider training set spectra with cluster membership probabilities comparable to AT2017gfo, and select the spectrum with the minimum Mahalanobis distance to AT2017gfo in the latent space.

At 1.4 days post-merger, we find that AT2017gfo most closely resembles a combination of Clusters 1, 3, and 5. Its latent space representation has 14% confidence of belonging to Cluster 1, 1% for Cluster 3, and 85% for Cluster 5, with negligible ≪1% probabilities for the other clusters. The training set spectrum with minimum Mahalanobis distance and comparable cluster membership to AT2017gfo is shown in Figure 9; it is mostly a blackbody shape, but it also has the observed 8000 Å Sr II absorption feature. Its physical parameters are: $\log_{10}(L_{\rm outer}/L_\odot) = 7.778$, $Y_e = 0.442$, $\log_{10}(\rho_0/{\rm g\,cm^{-3}}) = -13.5$ ($\rho_0 = 3 \cdot 10^{-14}\,{\rm g\,cm^{-3}}$), $v_{\rm inner}/c = 0.332$, $v_{\rm exp}/c = 0.229$, and $s/k_{\rm B} = 31.275$.

---

[12] Acquired through WISeREP (Yaron & Gal-Yam 2012).
[13] http://www.astroml.org/_modules/astroML/dimensionality/iterative_pca.html

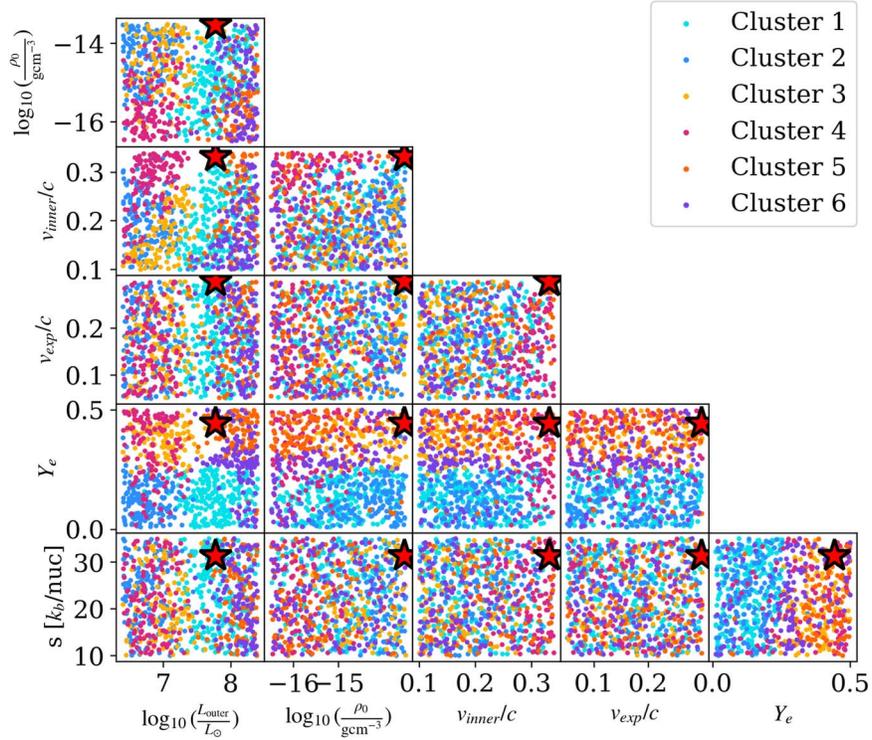

**Figure 8.** Corner plots for the 1.4 days post-merger spectra with at least 99% confidence of belonging to one cluster (colored by cluster). Each projection compares two of the available physical parameters used to generate the simulated kilonova spectra: $L_{outer}$, $\rho_0$, $v_{inner}/c$, $v_{exp}/c$, $Y_e$, and $s$. Clusters occupy distinct regions of physical parameter space; these regions are most apparent in the first column. The red star in each projection indicates the predicted location of confirmed kilonova AT2017gfo, based on the latent space position and cluster membership of its 1.4 days spectrum.

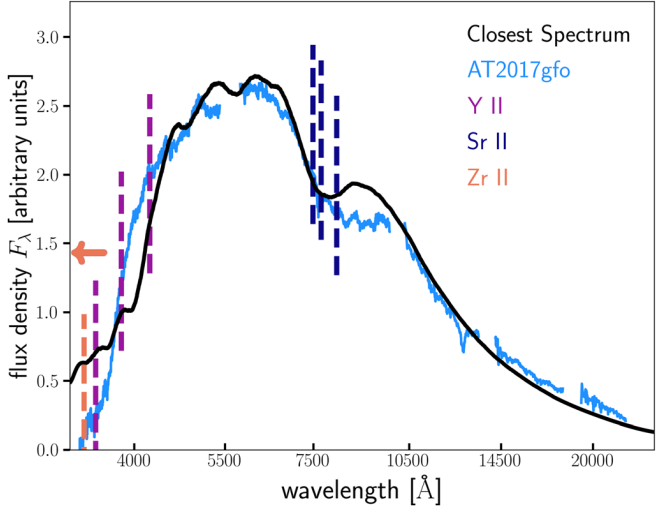

**Figure 9.** Comparison of the observed AT2017gfo spectrum at 1.5 days post-merger (blue line) with our closest data set spectrum (minimum Mahalanobis distance, comparable cluster membership, and most similar Sr II feature; black line). We mark several of the strongest absorption lines, which come from a combination of Sr II, Y II, and Zr II. The orange arrow indicates that one of Zr II's transition line centers is beyond the edge of the plot.

This is not an attempt at precise parameter estimation, but rather we wish to associate the spectrum to a general region of physical parameter space. AT2017gfo's proxy spectrum with the physical parameters provided above is marked with a red star in Figure 8. This point falls in a region of parameter space which appears sparsely populated, most clearly seen in the $v_{inner}/c$ and $Y_e$ rows of Figure 8's leftmost column. This sparse region is due to our choice to only display spectra with at least 99% confidence of belonging to one cluster, so that the clusters can be more easily seen. The points in the region near AT2017gfo are not displayed simply because they have significant probabilities of belonging to two or more clusters. The Cluster 3 and Cluster 5 points around the red star (and the sparse region) fall within $Y_e \gtrsim 0.3$, consistent with other works (Gillanders et al. 2022; Vieira et al. 2023a). Gillanders et al. (2022) also try fitting a spectrum with a high $Y_e$ of ∼0.44 to the AT2017gfo spectrum, and rule it out primarily because it produces a Ca II absorption feature that is not present in the observed spectrum. We do not see this Ca feature in our closest spectrum, which also has $Y_e$ ∼0.44. Assuming that our synthetic spectra are accurate representations of observed kilonovae spectra, we attribute the absorption at short wavelengths to Y II and Zr II, with the ∼8000 Å dip arising from Sr II (see Figure 9).

## 4. Discussion

### 4.1. Interpreting the Clusters' Dominant Absorption Lines

The presence of some combination of Y, Zr, and Sr at low wavelengths for many of the clusters is in agreement with the analyses of Gillanders et al. (2022) and Vieira et al. (2023a) for AT2017gfo. Watson et al. (2019) observe that these three elements should be the most easily detected of the first $r$-process peak elements, due to their low excitation potentials. Sr in particular is closest to the first $r$-process peak ($A \sim 80$). They also note that the dominant Sr II transition line triplet's atomic levels are metastable, and these lines may be significantly strengthened by photoexcitation. For all these reasons, Sr II's prominence in the spectra is unsurprising.

Table 4
Key Physical Properties and Dominant Absorbing Elements for Each Cluster.

| Cluster | $Y_e$ | $\log_{10}(L_{outer}/L_\odot)$ | $\log_{10}(\rho_0/\text{g cm}^{-3})$ | $v_{inner}/c$ | Dominant Ions[a] |
|---|---|---|---|---|---|
| 1 | Low | High | Moderate | Wide Range | Ce III, Tb III, Nd III, Pr III, Sr II |
| 2 | Low | Low | High | Moderate | Ce II, La II, Sr II, Zr I |
| 3 | High | Moderate | High | Moderate | Sr II, Y II, Zr II, Ca II |
| 4 | Very High | Moderate | Very Low | High | Sr II, Y II, Zr I, Ni I (redder)[b] |
| 5 | Very High | High | Very Low | Wide Range | Sr II, Y III, Zr II, Cr II (bluer)[b] |
| 6 | High | High | Wide Range | Low | Xe II, Ru II, Mo II, Zr II |

**Notes.** The spreads in entropy and expansion velocity are too large to discern a meaningful trend for any of the clusters.
[a] The top ∼four elements with the most prominent spectral absorption features. See Section 3.2.1 for further discussion of specific ions and transition lines.
[b] These clusters' spectra are mostly blackbodies with very little absorption.

Cluster 1 is the most lanthanide-rich, with the lowest $Y_e$, and its dominant species are the most ionized of all the clusters. This is explained by the fact that lanthanides are f-shell elements with lower excitation potentials than other r-process elements (Tanaka et al. 2020). In particular, the presence of Ce III in this Cluster's spectra echoes findings of other recent studies such as Domoto et al. (2022), Gillanders et al. (2023a), and Shingles et al. (2023).

In Cluster 3, there is a subset of spectra with absorption near 6500 Å. Based on Figure 7, we attribute this feature to a combination of Y II and Ca II. This is in agreement with Gillanders et al. (2022), who attribute this feature to the Ca II 3d–4p triplet. Domoto et al. (2021) also identify this triplet absorption feature in their higher $Y_e$, lanthanide-poor kilonova models. Sneppen & Watson (2023) also propose that the Y II feature (rest wavelength at ∼7800 Å, blueshifted to ∼6300 Å) is detectable in the 3+ days post-merger spectra of AT2017gfo. This further supports our findings that AT2017gfo closely resembles Cluster 3 spectra (see Section 3.4).

In Cluster 6, both Mo and Ru are likely formed from the weak r-process (Hansen et al. 2014). The cluster's Xe II identification is particularly interesting because Xe corresponds to the second r-process peak ($A \sim 130$; Xe has $A \sim 131$).

The two deepest absorption features in the clusters' spectra are due to Sr (Cluster 3) and Xe (Cluster 6), in spite of both elements having relatively low opacities (open s-shell and p-shell, respectively; see Tanaka et al. 2018, 2020 for more discussion on open shell opacities). Both elements, however, are abundantly produced because they are located near an r-process peak. This indicates that strong absorption can result from either high opacity or high abundance.

### 4.2. Connecting Absorption Features with Physical Parameters

Physical parameters including $Y_e$, $L_{outer}$, $\rho_0$, and $v_{inner}/c$ display varying degrees of correlation with the dominant absorption features in each cluster's spectra. We provide a qualitative comparison of the physical parameters associated with each cluster, as well as the elements with the most prominent spectral absorption features, in Table 4. A rigorous quantitative description of the correlations between the clusters and physical parameters is beyond the scope of this work, and should be pursued with a more physically sophisticated and realistic training set (see Section 4.4 for further details on how simulations could be improved).

In general, $Y_e$ and $L_{outer}$ are the parameters most closely tied to spectral absorption features. We find that lower $Y_e$ corresponds to heavier element production with strong lanthanide absorption features. At $Y_e \gtrsim 0.25$ (i.e., Clusters 3–6), we do not see any prominent lanthanide absorption features. At the highest $Y_e \gtrsim 0.35$ (Clusters 4 and 5), we see only faint absorption from first r-process peak elements, indicating that only the "weak" r-process occurs in these kilonova conditions.

$L_{outer}$ influences the location of the spectral peak, with higher luminosity shifting the peak to bluer wavelengths and increasing the peak amplitude (e.g., Cluster 2 compared to Clusters 5 or 6). Low $\rho_0$ seems to inhibit the formation of very deep absorption features—this is most apparent in Clusters 4 and 5, though the lack of absorption features can also be attributed to high $Y_e$. With a fixed outer ejecta boundary velocity, the photospheric velocity $v_{inner}/c$ predominantly determines how blueshifted the observed absorption features are from their rest wavelengths.

### 4.3. Comparison with Other Studies

We are aware of only one other study that applies dimensionality reduction to a kilonova spectra data set, called KilonovaNet (Lukošiute et al. 2022). The KilonovaNet tool trained a supervised conditional VAE on three spectral data sets (Kasen et al. 2017; Dietrich et al. 2020; Anand et al. 2021) and their corresponding input physical parameters. The latter two data sets are generated by the POSSIS radiative transfer code (Bulla 2019, 2023). Spectra in POSSIS are less computationally expensive to generate than in TARDIS because POSSIS takes opacities as a pre-computed input. This makes it difficult to link absorption features with individual ions and transition lines. Lukošiute et al. (2022) do not attempt to identify individual absorption features or determine trends of kilonovae spectra linked with certain physical properties. In contrast, TARDIS calculates opacities during the simulation based on the number density of each ion in the plasma (which is dependent on the element abundance pattern). TARDIS's configuration enables us to directly link spectral features with specific elements, ions, and transition lines, in addition to the physical parameters of the kilonova ejecta.

### 4.4. Limitations and Caveats

All analysis of observed spectra using our trained AE model is based on the assumption that our synthetic spectra accurately reflect kilonovae in nature. However, this study makes several simplifying assumptions in order to generate a training set of spectra. These assumptions limit our ability to fully reproduce and interpret real-life spectra. We describe some of the main limitations below.

While we successfully identified the dominant spectral absorption lines for the experimental line list described in Section 2.2, we caution that this list is lacking complete data for certain r-process species. We have purposefully excluded all theoretical lines to be more confident in our absorption line identifications. No line list—neither experimental nor theoretical—contains all absorbing species relevant to kilonova. Because of this limitation, there may be additional prominent lines that do not appear in our spectra. For example, very few lanthanide lines below ∼6000 Å have been experimentally measured, despite their predicted absorption contribution at short wavelengths (e.g., Tanaka et al. 2020). This could lead to overestimated emission at short wavelengths in simulated spectra; indeed, we see this mismatch below 4000 Å in our closest spectrum to AT2017gfo at 1.5 days post-merger (Figure 9), with Gillanders et al. (2022) and Vieira et al. (2023a) showing similar discrepancies. This does not, however, invalidate the experimentally verified absorption lines that we have already identified, and this work is not attempting to perfectly fit any observed spectrum.

TARDIS's assumptions of a 1D, spherically symmetric ejecta are an oversimplification and also impact the resulting kilonovae spectra. Multi-dimensional general relativistic magnetohydrodynamics simulations show that NS mergers may produce multiple asymmetrical ejecta components with different physical properties (e.g., Wanajo et al. 2014; Just et al. 2015; Mendoza-Temis et al. 2015; Wu et al. 2016; see Fernández & Metzger 2016; Radice et al. 2020 for reviews). A kilonova with multiple components may have a different appearance depending on the viewing angle and time (Kasen et al. 2015; Wollaeger et al. 2018; Darbha & Kasen 2020; Kawaguchi et al. 2020; Korobkin et al. 2021; Nativi et al. 2021; Shingles et al. 2023). Future work will simulate multiple radial ejecta shells to explore the impact of their different compositions and/or densities.

Additionally, TARDIS is time-independent and only simulates one epoch of kilonovae spectra at a time—we cannot evolve physical parameters between epochs. In reality, the luminosity should evolve over time based on the heating rate (Rosswog & Korobkin 2022) and thermalization efficiency (e.g., Barnes et al. 2016). This self-consistent luminosity evolution is necessary to produce realistic kilonova light curves, but in this work our focus is on identifying spectral absorption features. For this reason, we opt to keep the physical parameters fixed across epochs and generate separate spectra training sets at epochs corresponding to observations of AT2017gfo. In future work, our simulated spectra could be directly connected to light curves through self-consistently computing heating rates. In the meantime, we can still link the physical parameters inferred from kilonova light curves to specific spectra clusters, thus quickly identifying which photometrically detected kilonovae will be most interesting for spectroscopic follow-up.

### 5. Conclusion

In this work, we demonstrate that dimensionality reduction combined with clustering can extract key information on different types of kilonovae spectra. We apply an unsupervised AE to convert a high-dimensional simulated spectra data set spanning kilonova parameter space into a 5D latent space. By running a BGMM algorithm on this latent space representation of the data, we find that there are approximately six clusters or "types" of kilonovae spectra that stand out. Each cluster has its own associated spectral features, and using TARDIS we can distinguish the exact elements, ions, and transition lines that dominate the absorption features.

Our AE and clustering results for the 1.5 days post-merger training set indicate that future observed kilonovae spectra could display absorption by a combination of weak r-process species including Sr II, Y I–II, and Zr I–II. These types of kilonovae in our simulations are characterized by $Y_e \gtrsim 0.3$; at the highest $Y_e$ spectra approach a featureless blackbody shape, with the blackbody peak dependent on $v_{inner}/c$ and $L_{outer}$. Strong lanthanide contributions (especially from Ce II–III) are possible at low $Y_e \lesssim 0.25$. Xe II can also produce strong absorption for especially bright kilonovae with $\log_{10}(L_{outer}/L_\odot) \gtrsim 7.5$ ($L_{outer} \sim 10^{41}$ erg s$^{-1}$), $Y_e \sim 0.3$, and $v_{inner}/c \lesssim 0.3$.

The later 2.5 and 3.5 days post-merger data sets (details provided in the Appendix) show an evolution to fainter and redder spectra, but with generally similar absorption features and physical properties persisting between epochs. More lanthanides emerge at later times, and Ce in particular starts to produce deep absorption features in multiple clusters.

We input AT2017gfo's spectrum into the latent space to determine its cluster membership along with the dominant absorbing species and physical properties associated with the cluster spectra. At 1.4 days post-merger, we find that the spectrum closest to AT2017gfo (and with comparable cluster membership) displays the same Sr II absorption feature identified in previous works. Additionally, this closest spectrum falls into the $Y_e \gtrsim 0.3$ region of kilonova physical parameter space, consistent with other works such as Gillanders et al. (2022) and Vieira et al. (2023a).

The current spectra training sets are limited to 1D single component configurations; future work involves building more sophisticated training sets using multi-component kilonova models with self-consistent time evolution of the physical parameters. Our AE and clustering analysis framework can be applied with other simulated kilonova spectral data sets. The ultimate goal is to apply this analysis framework to future observed kilonovae spectra as we have done with AT2017gfo, to gain context on the dominant absorption features and region of physical parameter space occupied by new spectra. This analysis will be especially fruitful if inferred physical parameters from observed light curves can be linked to certain types of spectra identified in this work.

While this work includes several important simplifying assumptions, we still produce a set of simulated spectra with features echoing many previous studies, as well as the single observed kilonova AT2017gfo. This exploratory study leverages dimensionality reduction to pinpoint several combinations of r-process species and absorption features that form a "zoo" of possible kilonovae with different physical conditions. These diverse kilonovae identified through our analysis provide a useful snapshot of what to expect and how to interpret future kilonovae spectra discovered in upcoming LIGO–Virgo–KAGRA observing runs.

### Acknowledgments

N.M.F. is from O'ahu, one of the islands of Hawai'i, an indigenous space where the descendants of the original people are kānaka 'ōiwi / Native Hawaiian. N.M.F. is a visitor in Tiohtiá:ke/Mooniyang, also called Montréal, which lies on the unceded lands of the Kanien'kehá:ka. We are grateful to the traditional stewards of the lands, waters, and skies that we gather within. As we engage in astronomy on these lands, we respect, listen to, and make space for indigenous voices and ways of knowing the sky and Earth that sustain us all.

The authors thank the anonymous reviewer for providing valuable feedback. This research made use of resources from the Digital Research Alliance of Canada (https://alliancecan.ca/en) and Calcul Québec (https://www.calculquebec.ca/). We thank Nikolai Piskunov and Eric Stempels for help with the Vienna Atomic Line Database (VALD; http://vald.astro.uu.se/~vald/php/vald.php) data, operated at Uppsala University, the Institute of Astronomy RAS in Moscow, and the University of Vienna. We thank Shinya Wanajo for sharing their reaction network calculations. This research made use of TARDIS, a community-developed software package for spectral synthesis in supernovae (Kerzendorf & Sim 2014; https://docs.astropy.org/en/stable/Kerzendorf et al. 2023). The development of TARDIS received support from GitHub, the Google Summer of Code initiative, and from ESA's Summer of Code in Space program. TARDIS is a fiscally sponsored project of NumFOCUS. TARDIS makes extensive use of Astropy and Pyne. N.M.F., N.V., J.J.R., and D.H. acknowledge funding from the Natural Sciences and Engineering Research Council of Canada (NSERC) Discovery Grant program and the Canada Research Chairs (CRC) program. N.M.F. acknowledges funding from the Fondes de Recherche Nature et Technologies (FRQNT) Doctoral research scholarship. N.M.F., N.V. and D.H. acknowledge support from the Canadian New Frontiers in Research Fund (NFRF)—Explorations program and the Trottier Space Institute at McGill. D.H. thanks the Institut de Planétologie et d'Astrophysique de Grenoble (IPAG) for supporting an extended visit during which this work was completed. N.V. acknowledges funding from the National Sciences and Engineering Research Council of Canada, the Murata Family Fellowship, and the Bob Wares Science Innovation Prospectors Fund. J.J.R. acknowledges funding from the FRQNT Nouveaux Chercheurs Grant program, Canada Foundation for Innovation, and the Québec Ministère de l'Économie et de l'Innovation.

*Software:* astropy: Astropy Collaboration et al. (2013; https://docs.astropy.org/en/stable/); cmasher: van der Velden (2020; https://cmasher.readthedocs.io/); hdbscan: McInnes et al. (2017; https://hdbscan.readthedocs.io/en/latest/index.html); pytorch: Paszke et al. (2019; https://pytorch.org/); scikit-learn: Pedregosa et al. (2011; https://scikit-learn.org/stable/index.html); TARDIS: Kerzendorf & Sim (2014; https://tardis-sn.github.io/tardis/index.html); UMAP: McInnes et al. (2018; https://umap-learn.readthedocs.io/en/latest/)

## Appendix
## 2.4 and 3.4 Days Post-merger Epochs

The same general set of clusters identified in the 1.5 days post-merger epoch persist at later epochs, though the BGMM weights are sometimes reordered. The spectra and physical parameters associated with each cluster are shown in Figures 10 and 12, respectively. The spectra shift to become redder and fainter, as expected due to the time evolution of luminosity and blackbody temperature. Figure 11 shows SDEC plots for Clusters 4 and 6 at 3.4 days post-merger (see Section 3.2.1 for an explanation of the SDEC analysis). These clusters highlight an increased presence of neutral to doubly ionized lanthanide absorption features at later epochs (e.g., Tb, Nd, Pr, Ce, dysprosium $_{66}$Dy, europium $_{63}$Eu, and samarium $_{62}$Sm), which agrees with other simulations (e.g., Domoto et al. 2022). At 3.4 days post-merger, the actinide thorium $_{90}$Th contributes some short-wavelength absorption in Cluster 4. This cluster has $Y_e \lesssim 0.25$ (see Figure 12), in agreement with the assessment of Holmbeck et al. 2019 that Th can be produced in low $Y_e$ ejecta. Spectra still display absorption from singly to doubly ionized Y, Zr, Sr (including the $\sim 8000$ Å Sr II triplet feature), and/or Xe at low wavelengths.

The BGMM struggles to identify clusters as confidently at 2.5 days post-merger, so the threshold weight for robust cluster identification is reduced from 0.090 to 0.080; in this configuration, the algorithm identifies essentially the same six clusters that appear at 1.5 days post-merger. At 3.4 days post-merger, seven clusters are identified as optimal by the BGMM (see Figure 10, right panel), though the additional cluster seems to be another blackbody with a luminosity (or temperature) somewhere between the original two identified blackbody clusters.

At both the later epochs, the issue of misidentifying spectra belonging to Cluster 6 is exacerbated in comparison to the 1.4 days post-merger training set. There also appears to be a subset of spectra in Cluster 3 for the 2.4 days post-merger training set with a deep absorption feature that more closely resembles the shape of Cluster 5. We attribute this to the clusters overlapping in latent space, leading to occasional mislabelling. The overall appearance of each cluster remains consistent across epochs.

The closest matches to AT2017gfo at these later epochs are shown in Figure 13. Each spectrum is shown normalized by its total flux in the specified wavelength range. In both epochs, the dominant absorption feature is still attributed to Sr II, with weaker, short-wavelength contributions from a combination of Y II, Zr II, and Ni I. Compared to 1.5 days post-merger, several of the kilonova physical parameters decrease a day later: $\log_{10}(L_{outer}/L_\odot) = 7.318$ ($L_{outer} \sim 8 \times 10^{40}$ ergs s$^{-1}$), $Y_e = 0.331$, $\log_{10}(\rho_0/\text{g cm}^{-3}) = -15.091$ ($\rho_0 \sim 8 \times 10^{-16}$ g cm$^{-3}$), $v_{inner}/c = 0.223$, $v_{exp}/c = 0.118$, and $s/k_B = 19.2$. This is generally in agreement with the findings of Gillanders et al. (2022), although their fitted $Y_e$ is somewhat lower ($\sim 0.3$). At 3.5 days post-merger, we find: $\log_{10}(L_{outer}/L_\odot) = 7.754$, $Y_e = 0.434$, $\log_{10}(\rho_0/\text{g cm}^{-3}) = -13.959$, $v_{inner}/c = 0.277$, $v_{exp}/c = 0.182$, and $s/k_B = 25.4$. We cannot estimate the uncertainties on these parameters with our approach, but based on the discrepancies between the observed 3.4 days post-merger spectrum and the closest training set spectrum, the uncertainties may be substantial at this epoch. The accuracy of the luminosity parameter value may be inhibited by training on flux-normalized spectra; this normalization leads to some loss of information on the relative scales of the spectra (i.e., the flux or luminosity).

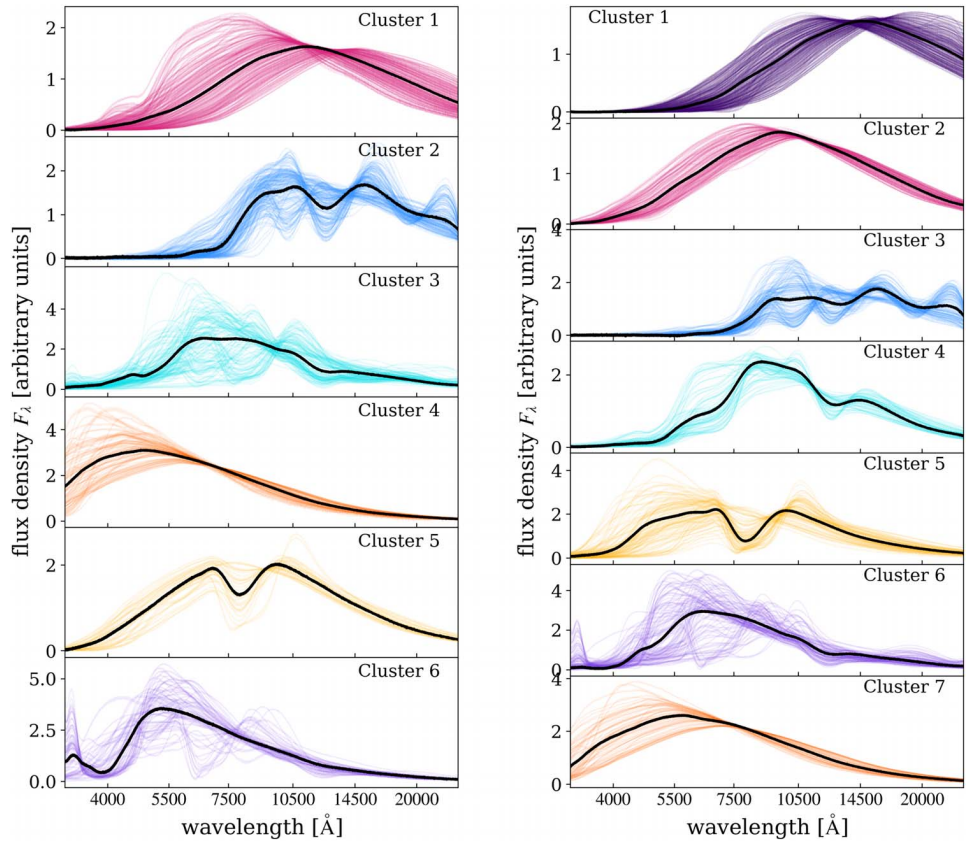

**Figure 10.** The 2.4 (left panel) and 3.4 (right panel) days post-merger spectra identified with at least 99% confidence to belong to one cluster, stacked and colored by cluster. The black line represents the mean spectrum for that cluster. In general, the cluster spectral shapes identified at 1.5 days post-merger and shown in Figure 6 persist in later epochs.

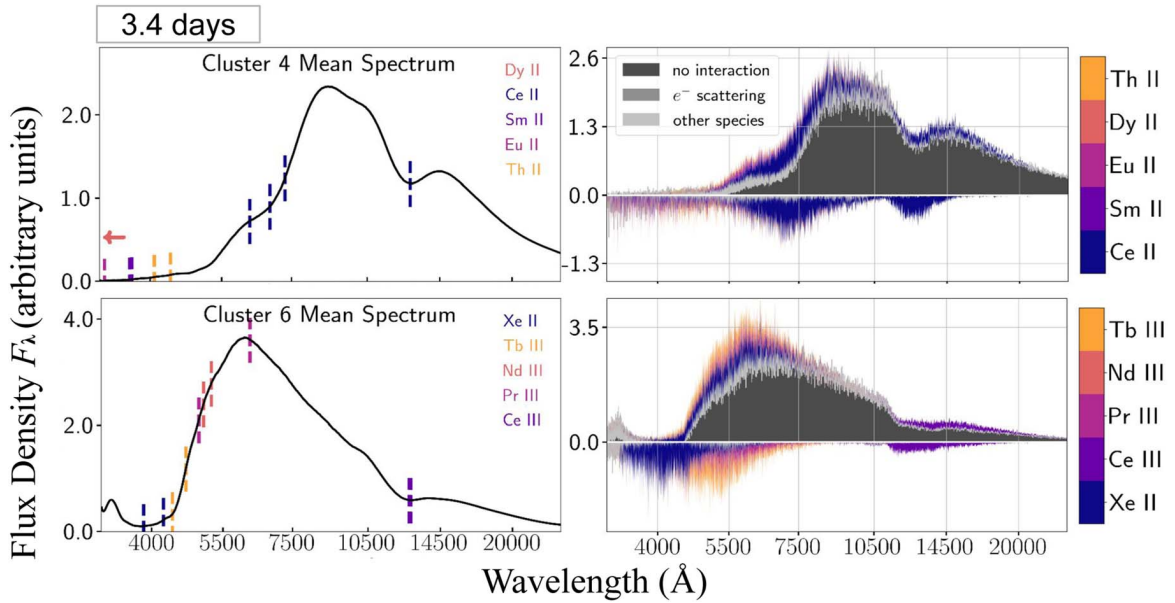

**Figure 11.** TARDIS Spectral element DEComposition (SDEC) plots for Clusters 4 (top row) and 6 (bottom row) at 3.4 days post-merger. Right column: in color are the absorption (below 0) and emission (above 0) for the top five contributing elements sorted by atomic number. Left column: vertical dashes mark the dominant absorption lines at their Doppler-shifted wavelengths; line centers beyond the edge of the plot are denoted with an arrow. An element may be assigned a different color based on its atomic number relative to the other plotted elements for a given cluster. Both clusters display more pronounced lanthanide features than at 1.4 days post-merger.

In both later epochs, the closest spectrum shows noticeable deviations from the AT2017gfo spectrum, especially in the position and shape of the main ∼8000 Å absorption feature. These issues may be alleviated with the incorporation of multiple kilonova components (Section 4.4), as seen in Vieira et al. (2023b). Additional limitations discussed in Section 4.4 may also impact our ability to accurately reproduce the absorption features of these observed spectra.

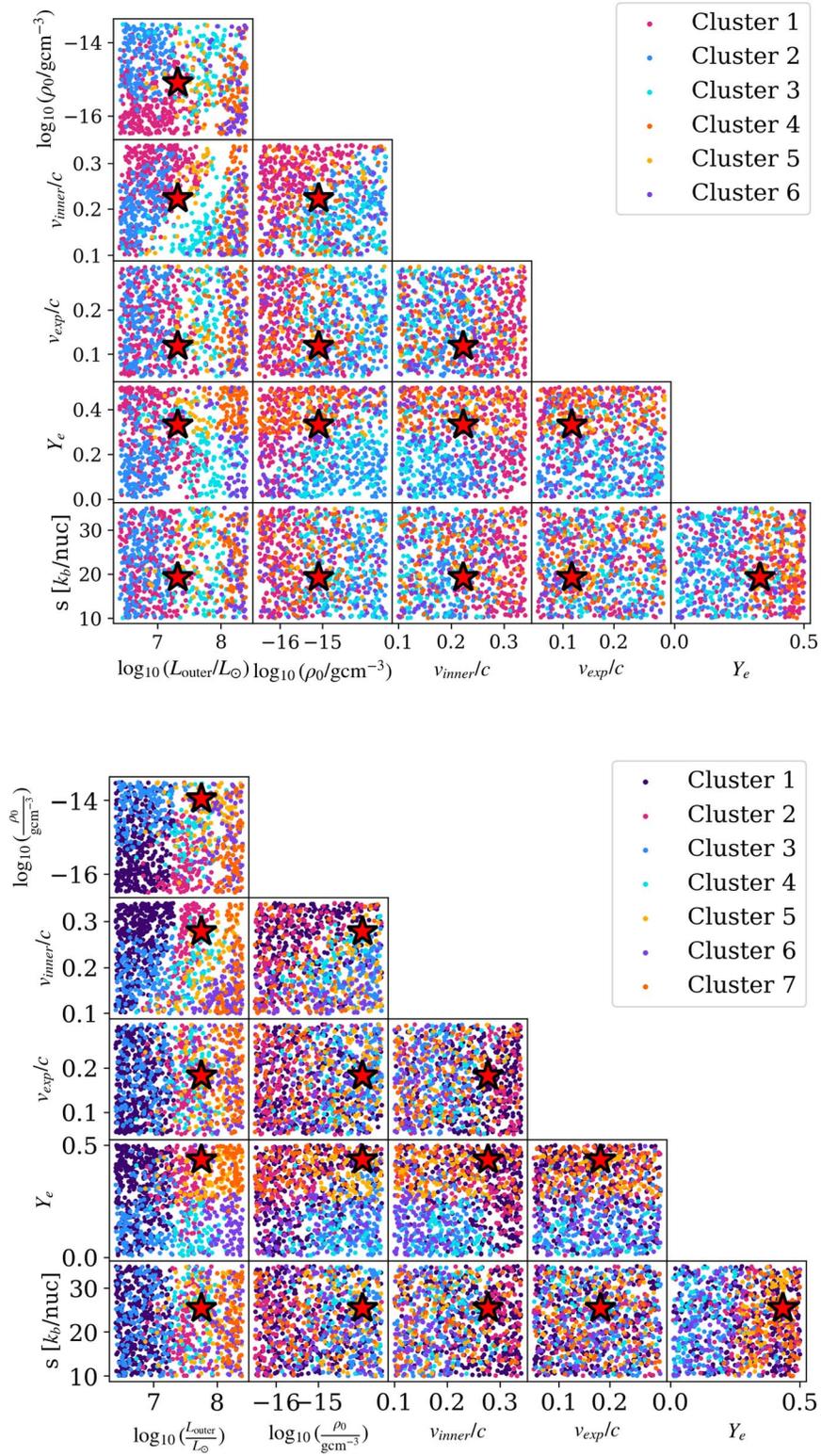

**Figure 12.** Corner plots for the 2.4 (top panel) and 3.4 (bottom panel) days post-merger spectra with at least 99% confidence of belonging to one cluster (colored by cluster). Each plot compares two of the available physical parameters used to generate the simulated spectra: $L_{\mathrm{outer}}$, $\rho_0$, $v_{\mathrm{inner}}/c$, $v_{\mathrm{exp}}/c$, $Y_e$, and $s$. There are distinct regions of physical parameter space occupied by individual clusters; these regions are most apparent in the leftmost column of plots. The red star in each plot indicates the predicted location of confirmed kilonova AT2017gfo, based on the position and cluster membership of its spectrum in latent space.

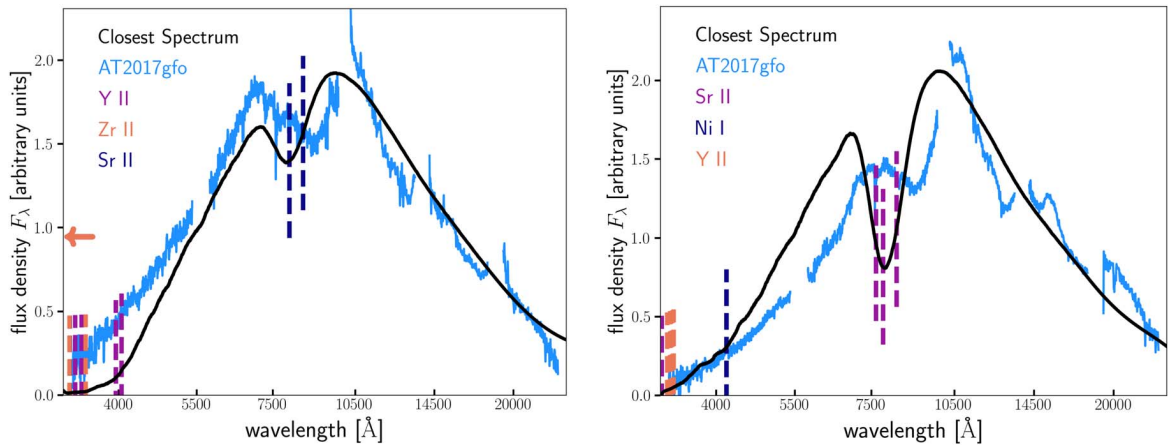

**Figure 13.** Comparison of the observed AT2017gfo spectrum (blue line) at 2.4 days (left panel) and 3.4 days (right panel) post-merger with our closest data set spectrum (minimum Mahalanobis distance, comparable cluster membership, and most similar Sr II feature; black line). We mark several of the strongest absorption lines, which come from a combination of Sr II, Y II, and Zr II (line center is off the edge of the plot, marked with an arrow). The closest training set spectrum bears less resemblance to the observed spectrum as we progress in time, but the general absorption features are still present.

## ORCID iDs

N. M. Ford ● https://orcid.org/0000-0001-8921-3624
Nicholas Vieira ● https://orcid.org/0000-0001-7815-7604
John J. Ruan ● https://orcid.org/0000-0001-8665-5523
Daryl Haggard ● https://orcid.org/0000-0001-6803-2138